\documentclass[aip,numerical]{revtex4-1}

\usepackage{graphicx}% Include figure files
\usepackage{dcolumn}% Align table columns on decimal point
\usepackage{bm}% bold math
\usepackage{amsfonts}
\usepackage{amsmath}
\usepackage{amssymb}
\usepackage{amstext}
\usepackage{latexsym}
\usepackage{textcomp}
\usepackage{color}
\usepackage{tikz}
\usepackage{natbib}


\begin{document}
\def\myfrac#1#2{\frac{\displaystyle #1}{\displaystyle #2}}

\title{Studies on ultra-short pulsed laser shock peening of stainless-steel in different confinement media}

\author{Kishore Elango}
\affiliation{Ruhr-Universit\"at Bochum, Chair of Applied Laser Technology,
Universit\"atsstra\ss e~150, 44801 Bochum, Germany}

\author{Jan S. Hoppius}
\email[Corresponding author: ]{hoppius@lat.rub.de}
\affiliation{Ruhr-Universit\"at Bochum, Chair of Applied Laser Technology,
Universit\"atsstra\ss e~150, 44801 Bochum, Germany}

\author{Lalit M. Kukreja}
%\email[Corresponding author: ]{hoppius@lat.rub.de}
\affiliation{Epi-Knowledge Foundation, C2 -- 4/1:3, Sector-4, Vashi, Navi Mumbai 400 703, India}

\author{Andreas Ostendorf}
%\email[Corresponding author: ]{andreas.ostendorf@ruhr-uni-bochum.de}
\affiliation{Ruhr-Universit\"at Bochum, Chair of Applied Laser Technology,
Universit\"atsstra\ss e~150, 44801 Bochum, Germany}

\author{Evgeny L. Gurevich}
\email[Corresponding author: ]{gurevich@lat.rub.de}
\affiliation{Ruhr-Universit\"at Bochum, Chair of Applied Laser Technology,
Universit\"atsstra\ss e~150, 44801 Bochum, Germany}

\date{\today}

\begin{abstract}

Laser Shock Peening (LSP) of stainless-steel 316 and 316L was studied using Ti: Sapphire laser pulses of about 2 ps duration, maximum energy of about 1mJ and pulse repetition rate of 5kHz in different liquid confinement media of Ethanol, Deionized water and separate aqueous solutions of NaCl and Glycerol. It is found that the laser fluence and/or energy attenuating mechanisms like self-focusing, filamentation, plasma breakdown in the confinement media are less significant with ps laser pulses than those with sub-ps or fs pulse durations. It is shown that the resulting surface hardness of the peened steel as a function of laser fluence depends significantly on the confinement media and the relative increase in the hardness increases monotonically with the acoustic impedance of the liquid of the confinement medium used during LSP.

\end{abstract}

%\pacs{\textit{79.20.Eb,73.20.Mf,81.16.Rf}} 

%\end{frontmatter}

\maketitle

\section{Introduction}

Laser shock peening (LSP) in the ultra-short pulse durations regime of pico-second (ps) to femto-second (fs) is a potentially important area of research for surface enhancement of engineering components of different metals and alloys even though this area is still in its infancy. This is because in this time regime the pressure of the shock waves generated is correspondingly orders of magnitude higher than that generated for example in nano-second or micro-second time regimes and the depth of the residual compressive strength is much finer \cite{Ageev2016, Hoppius2018}. Also, the process of ultrashort pulse shock peening is fundamentally non-thermal in nature because the laser pulse duration in this case is comparable or shorter than the electron-phonon coupling time. The heating and cooling rates of the surface in this case can be as large as tera Kelvin per second \cite{Duff2007} resulting in ultra-fast ablation, which is not possible with longer laser pulses resulting in slower thermal processes like melting and vaporization. The end result of these initial processes is the generation of sharp shock-wave \cite{Noack1998}, which leaves residual compressive stress in the surface region of the subjected metals or alloys causing its peening through plastic deformation \cite{Lee2011}. 

Mostly LSP is carried out by applying a sacrificial layer on the metal or alloy surface and an overlayer of confinement medium, generally water \cite{Fairand1976, Fox1974}. Even though some researchers prefer to carry out LSP in fs time regime without a sacrificial layer \cite{Sano2006}, which we have found to result in compromised surface quality \cite{Hoppius2018}. But even in the former case a confinement layer is invariably used because the amount of pressure created due to laser ablation under the confinement medium is several times or even an order of magnitude larger than the pressure created in an unconfined open environment \cite{Fabbro1998}. When the high intensity laser pulse interacts with the target material, a high recoil pressure ablation plume is generated and in the absence of a confinement layer this plume expands and loses most of its energy in the surrounding environment away from the target surface rather quickly. Whereas, in the presence of a confinement medium the expanding plume is confined and thus, the shock energy is concentrated more towards the target surface for longer time, which is beneficial in peening the target material by resulting in enhanced compressive residual stress on its surface region \cite{Hong1998}. It has been found that the pressure induced by medium-confined ablation plume can be as high as an order of magnitude higher and 2 to 3 times longer than pressure induced by the same plume in air or vacuum, for the same laser intensity \cite{Fabbro1990}. It is therefore imperative to choose a good confinement medium to increase the peening efficiency. 

Choice of a good confinement medium depends on its optical transparency for the laser wavelength, chemical inertness to sample material and sacrificial layer, high acoustic impedance and ease of usability. The layer of the confinement medium should also have an optimum thickness. If the confinement layer is very thin, it makes the processing difficult particularly at high laser intensities because it remains ineffective for confinement of the ablation plume produced. And if the confinement layer is too thick, it absorbs, distorts the pulsed laser beam that passes through this layer and attenuates the incident laser energy due to filamentation particularly at shorter pulse durations and higher intensities \cite{Hoppius2019}. That is why it is also advantageous to take a confinement medium with smaller linear and nonlinear refractive indexes $n_0$ and $n_2$ to increase the self-focusing threshold \cite{Couairon2007}.

To the best of our knowledge, N. C. Anderholm was the first to present the technique of confined laser produced ablation \cite{Anderholm1970}. In this study he investigated a method to generate laser induced stress waves under a transparent overlayer of quartz disk. The experiments were carried out using a ruby laser system at pulse energy of about 7~J and duration of 12~ns. Subsequently several studies have been carried out, which showed increased efficiency of the peening with confinement layer. D. Devaux et al. investigated the characteristics of laser-induced shock wave in a confining environment \cite{Devaux1991}. Experiments were carried out with a conventional short ns time regime pulsed neodymium glass laser at laser intensities ranging from 107 to 1011 $W/cm^2$. Laser-induced shock pressures were measured using a synthetic piezoelectric quartz gauge under glass, water and without any confinement layer. They found that the shock pressure generated due to the laser ablated plume / plasma under a confinement layer depends on the square root of the combined acoustic impedance of the confinement medium and the target material \cite{Devaux1991}. They observed the maximum increase in the shock pressure by about an order of magnitude due to the confinement layer in a broad range of the laser intensity \cite{Devaux1991}. 

Followed by these studies numerous researchers have studied the conventional LSP under different confinement media in long and short pulse duration regimes. However, to the best of our knowledge, a systematic study on the influence of different confinement media used during LSP in ultra-short laser pulse duration regime of ps - fs is not yet reported in the pertinent literature.  We carried out such a study on the role of different confinement media on the mechanical properties of stainless steel subjected to LSP in the ps time regime. We choose four confinement media for this study viz. Ethanol, Deionized water, aqueous solutions of NaCl and Glycerol separately.  Since pure glycerol is highly viscous, the confinement media in this case was prepared by mixing water in it so that we could flow it over the target sample to remove the ablation debris and bubbles etc. at the high repetition rate of the peening laser pulses. Using these confinement media for ps LSP we first investigated the problems of laser induced non-linear effects like self-focusing, filamentation and plasma breakdown etc. in the confinement media, which attenuate the laser energy that can be delivered to the surface of the sacrificial layer. Followed by that, under the optimized conditions of LSP, we studied the surface hardness of the steel subjected to ps LSP under different confinement media. The results of these studies are presented and discussed in this paper.

\section{experimental}

\begin{figure}
\begin{center}
\includegraphics[width=8cm]{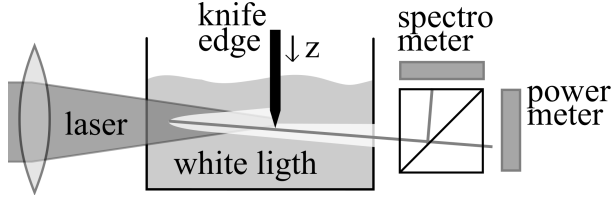}
\end{center}
\caption{Schematic of experimental arrangement for the measurements of the incident beam attenuation due to filamentation in different confinement media.}
\label{setup1}
\end{figure}

This study was carried out in two parts. Part one was to study the laser energy attenuating processes at ultra-short durations of the sub-ps and ps of the laser pulses used for the LSP such as non-linear effects of self-focusing, plasma breakdown and filamentation in different confinement media. Schematic of the experimental set-up for investigating the energy attenuation due to filamentation in different confinement media is shown in figure~\ref{setup1}. The ultrashort laser pulses of duration of $\sim$ 100~fs to 5~ps, wavelength of 800~nm $\pm$30~nm and pulse energy of about 1~mJ was focused by a plano-convex spherical lens of focal length of $\sim$ 100~mm into a test cuvette made of BK7 glass containing the confining liquid under investigation. The focal spot size of the laser beam in water, determined by ablation experiments was $\sim$ 20$\mu$m. A sharp metal blade was inserted incrementally into the beam path from top to bottom of the cuvette in the focal plane of the laser beam and after the light filament and white-light generation began as shown in Fig.~\ref{setup1} schematically, blocking the initial laser radiation and white light in small steps. The out coming laser and white lights were split to simultaneously pass into a power meter and a spectrum analyzer ({\it USB2000+} made by {\it Ocean Optics}). This combined and simultaneous measurements of laser and white light power and spectrum enabled the differentiation between white light and Ti:Sapphire laser radiation through computational deconvolution and acquiring the spatial intensity distributions of the two lights at different pulse durations, as we will discuss in the section \textit{Results and Discussions}.

\begin{figure}
\begin{center}
\includegraphics[width=8cm]{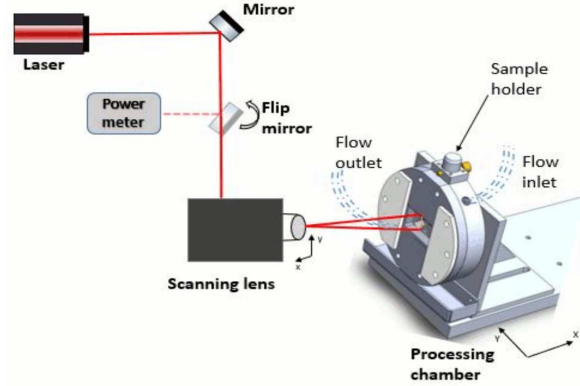}
\end{center}
\caption{Experimental set-up for studying the influence of different confinement media on ultra-short pulsed laser shock peening.}
\label{setup2}
\end{figure}

Part two of our study was to find the effect of different confinement media on the ultra-short pulsed laser shock peening of stainless steel 316 and 316L. The experimental set-up for studying this part is shown in figure~\ref{setup2}. We used a Ti:sapphire laser system ({\it Spitfire Ace} produced by {\it Spectra Physics}) operating at wavelength about 800 nm. The pulse duration was stretched by adjusting the compressor of the laser system to about 2~ps and the maximum peak energy was taken about 1~mJ at 5~kHz pulse repetition frequency. As shown in figure~\ref{setup2}, the laser beam was focused using a telecentric lens of about 128 mm focal length to the spot diameter of about 10 $\mu$m on to the target samples through the glass window of an in-house fabricated processing chamber and the liquid confinement media therein. The target samples were mounted in this confinement media filled processing chamber. Liquids of confinement media were pumped into the processing chamber using an pump to achieve a laminar flow to remove any bubble formed due to the repeated interaction of the laser beam with the target under the confinement media and the ablated debris that would invariably affect the parameters of the laser light adversely. This chamber (volume of $\sim$ 20 cm$^3$) was flushed with different liquids of the confinement media flowing at a rate of about 4.2~l/min for the aforesaid purpose. The chamber was mounted onto an X, Y-stage for positioning the samples. During the LSP process, a galvo-scanner ({\it SCANcube10}, manufactured by {\it SCANLAB}) directed the laser beam on to the target as shown in figure~\ref{setup2}.

Processing of the samples was carried out at different laser fluences controlled by a $\lambda$/2 wave plate and a polarizing beam splitter. The average laser power was measured prior to the processing using a power meter aligned in the laser beam path as shown in figure~\ref{setup2}. As mentioned, we carried out the LSP of stainless steel 316 and 316L under different confinement media of Ethanol, Deionized water, aqueous solutions of NaCl and Glycerol separately using a sacrificial layer of a black vinyl adhesive tape. Vinyl tape satisfies all the requirements of a good sacrificial layer such as good absorption of laser light, an even surface, chemically inert, good sticking to the target metal, environmentally safe etc., which makes it a better alternative in comparison to layers like black paint, transparent adhesive tape etc. The laser peened samples were investigated for the resulting surface quality using a scanning electron microscope, for chemical analysis using Energy-dispersive X-ray spectroscopy (EDX) and surface hardness using a Vickers hardness tester ({\it KB30BV7}) with HV 0.1 (equivalent to 0.981N).

\section{Results and Discussions}

\begin{figure}
\begin{center}
\includegraphics[width=8cm]{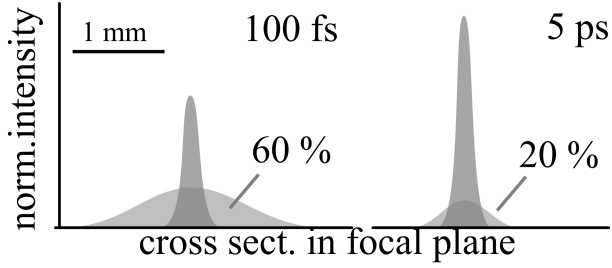}
\end{center}
\caption{Comparative spatial intensity distributions of the incident laser radiation (dark-gray sharp peak) and of the white light generated through filamentation (light-gray broad peak) in cases of 100 fs and 5 ps laser pulses.}
\label{IntDistr}
\end{figure}

As stated earlier, we first investigated the laser pulse energy attenuation processes due to combined thermal and non-linear effects of the laser beam passing through the confinement media used for the peening. Using the experimental setup shown in figure~\ref{setup1}, a cross sectional spatial intensity distribution of the laser light and the white light generated through filamentation, which attenuated the laser energy and fluence, were acquired. These spatial intensity profiles of the two lights are shown in figure~\ref{IntDistr}. For 800 $\mu$J pulses, transmitting through 40mm length of water from the focal spot of the laser light, two important inferences were elicited. 1) The focal spot diameter of the Ti:Sapphire radiation increased by a factor of almost 7.5 to about $\sim$150 $\mu$m reducing the fluence at the target correspondingly. 2) Depending on the pulse length, as shown in figure~\ref{IntDistr}, large portion of the laser radiation was found to get converted into white light. The pulse energy loss due to this filamentation and white light generation process was found to be about $\sim$60\% for the 100~fs pulses while it was only about 20\% in case of the laser pulse duration of 5 ps, as shown in figure~\ref{IntDistr}. From this point of view, it is energetically advantageous to use a few ps laser pulses compared to that of 100 fs for shock-wave peening in a confinement medium. Since the durations of the 100~fs and 2-5~ps pulses do not exceed the electron-phonon coupling time in metals, we assume that the amplitudes of the shock waves generated with these pulses are comparable at the constant laser fluence delivered at the target metal. Hence it is imperative to investigate the relative merits and demerits of LSP in these two duration regimes of the laser pulses.

\begin{figure}
\begin{center}
\includegraphics[width=16cm]{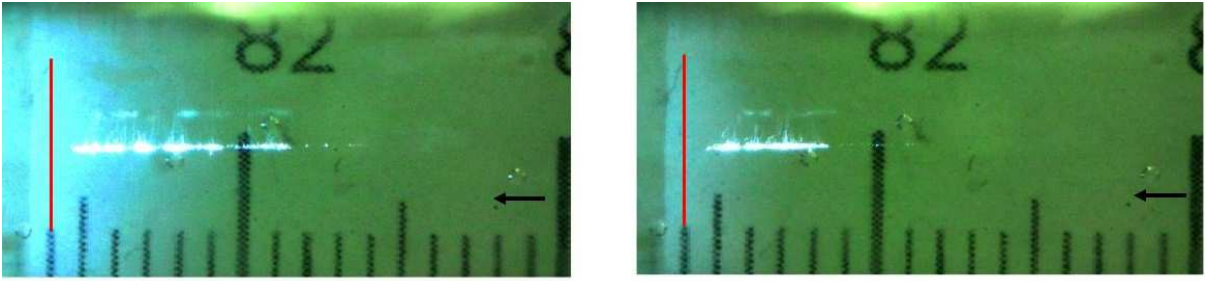}
\end{center}
\caption{Longitudinal size of filamentation and its point of initiation observed in deionized water as a confinement medium at laser pulse duration of (a) 100 fs (b) 2 ps. The beam propagation direction is indicated by black arrows. The red vertical line shows the focal point of the focusing lens in the respective image.}
\label{filaments}
\end{figure}

One of the most important issues is the optimization of the confining media thickness. Figure~\ref{filaments} shows a comparison of the length of the onset and the actual extension of the filamentation produced in deionized water as confinement medium for the laser pulse duration of 100~fs and 2~ps at identical pulse energy of about 1~mJ. As can be seen filamentation and plasma breakdown initiate much later along the path of the laser beam in the deionized water in case of the 2~ps than 100~fs. A similar behavior was also observed in ethanol \cite{Hoppius2019}. This gives a bench mark of the length of the confinement medium, which should be used to locate the target sample so as to zap the laser beam with minimal loss of energy.

In our experimental studies we have also investigated the surface morphologies and contamination mainly due to oxidation and carbonization. Our general finding was that there were no significant differences in these aspects due to the different confinement media as long as the laser beam interacted with the sacrificial layer and it was not allowed to interact with the metal surface directly. As stated earlier, different confinement media used in this study were Ethanol, Deionized water, aqueous solution of NaCl at 20\% wt./v and glycerol at 60\% concentration also in water. The pertinent physical properties of these confinement media are summarized in table~\ref{Tab1}. Deionized water was preferred over the normal water, to avoid any mineral deposition in the chamber and on the samples usually caused by the normal water. To study the effect of different confinement media we used two different samples of steel and measured the surface hardness on Vickers scale at different fluences of the 2 ps laser pulses used for the peening. The reference hardness in the following figures refers to the hardness of the untreated steel samples and the experimental points are the hardness relative to this reference. Throughout this work untreated reference hardness (HV) value was taken as 160 HV for grade SUS316L and 290 HV for SUS316, both measured using Vickers hardness testing equipment. In these experiments the sample surface was covered by overlapping laser pulses. The repetition rate of the laser was 5 kHz, the scanning velocity 10~mm/s, corresponding to 2~micrometer shift between the centers of two consecutive pulses.

\begin{table}
 \caption{Pertinent properties of different confinement media used in this study, $^*$ are taken from \cite{R1} and $^{**}$ are from \cite{Negadi2017}.}
 \label{Tab1}
 \begin{tabular}{|c|c|c|c|}
\hline
Confinement layer     & Density  & Sound speed & Acoustic impedance\\
                      &(kg/m$^3$)& (m/s)       &   10$^6$ (kg m$^{-2}$s$^{-1}$)\\
\hline
Ethanol$^*$           & 789      & 1180        &  0.93  \\
Deionized water$^*$   & 998      & 1400        & 1.48   \\
20\% NaCl$^*$         & 1150     & 1750        & 1.57  \\
60\% Glycerol$^{**}$  & 1230     & 1899        & 2.34  \\
\hline
\end{tabular} \\
\end{table}

\begin{figure}
\begin{center}
\includegraphics[width=16cm]{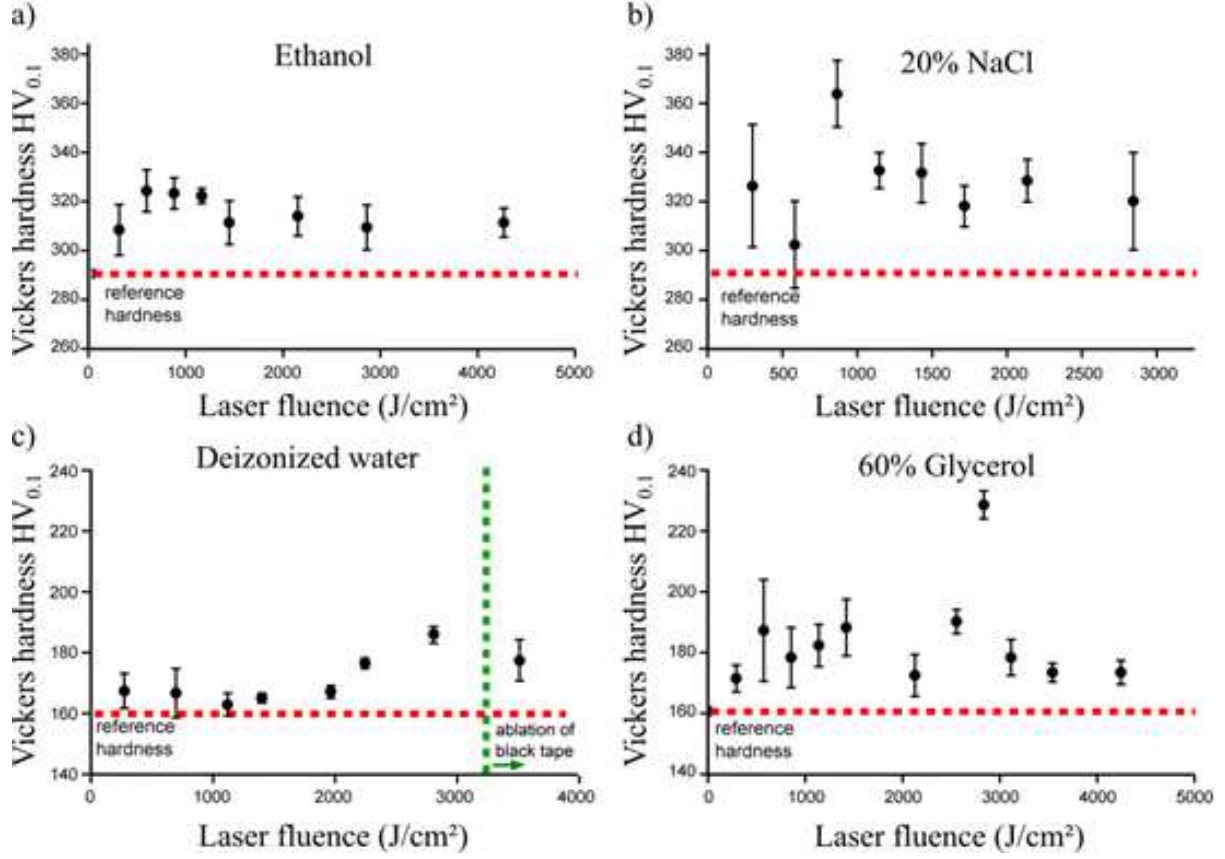}
\end{center}
\caption{(a-d). Variation in the surface hardness of stainless steel SUS316L peened at different fluences of the laser pulses in (a) Ethanol (b) Deionized water (c) aqueous solution of NaCl of 20\% concentration and (d) 60\% Glycerol in water as confinement media.}
\label{AllTogether}
\end{figure}

Figure~\ref{AllTogether}(a) shows the hardness of the stainless steel SUS316 measured at different fluences of the laser beam in the confinement medium of Ethanol. On the x-axis we plot calculated fluence, which is the pulse energy measured before the processing chamber divided by the measured laser spot area.  As can be seen in this figure the surface hardness briefly increased with laser fluence and then monotonically decreased before almost saturating at the laser fluence of about 1500~$J/cm^2$. The initial increase in the hardness is plausibly due to increase in the shock pressure induced by the increasing fluence of the peening laser pulses. Subsequent decrease in the hardness can be attributed to the onset of energy attenuating processes due to non-linear absorption, filamentation etc.

Experimentally measured hardness of the SUS316L stainless-steel as a function of the peening laser fluence in the confinement medium of Deionized water is shown in figure~\ref{AllTogether}(b). In contrast to the case of Ethanol stated above, in Deionized water, the surface hardness shows a little decrease in its value with increasing fluence in the beginning before a drastic increase in the hardness with the fluence preceding the final fall due to onset of the complete ablation of the sacrificial layer on the surface of the target sample. Complete ablation of the sacrificial layer at the laser fluence of about 3250~$J/cm^2$ was confirmed by visual inspection of the sacrificial layer and onset of Laser Induced Periodic Surface Structures (LIPSS) on the target metal observed under high magnification of SEM. Earlier we had shown that followed by the complete ablation of the sacrificial layer, whenever there is direct interaction between the laser pulses and the metal surface, LIPSS formation is inevitable \cite{Hoppius2018}. Formation of LIPSS, oxidation of the exposed surface of the target metal and partial ablation at the surface collectively seem to be responsible for the drastic decrease in hardness beyond the laser fluence of 3250~$J/cm^2$, where complete ablation of the sacrificial black tape occurred. Reason for the initial gentle decrease in the hardness is not clear to us at this stage and this could also be due to the fluctuations in the experimental conditions and/or errors in the measurements.

Dependence of the surface hardness on laser fluence in the confinement medium of aqueous solution of 20\% NaCl is shown in figure~\ref{AllTogether}(c). This dependence and the one in the confinement medium of 60\% glycerol in water, as shown in figure~\ref{AllTogether}(d) are rather similar in trend. First the hardness increases with the fluence due to the increase in the shock pressure generated and beyond the maxima, it decreases putatively due to the onset of laser energy attenuating processes as discussed earlier. However, an interesting feature in figure~\ref{AllTogether}(d) is the spike of increases in the hardness in the fluence range of 2000~-~3500~$J/cm^2$ with the exceptionally high peak hardness of about 230 on the Vickers scale. This can possibly be explained by the following mechanism: The linear refractive index $n_0$ and the nonlinear refractive index $n_2$ for glycerol are both larger than that of deionized water \cite{Brown1979}, hence, their product in the water-glycerol mixture is also larger than that in the water. The critical laser pulse power needed for filamentation in transparent media (as shown in figure~\ref{filaments}) can be estimated \cite{Couairon2007} as $P_{cr}=\frac{0.15 \lambda^2}{n_0 n_2}$, where $\lambda$ is the incident light wavelength. Moreover, the ionization potential of glycerol is lower that of water (10.1~eV and 12.6~eV respectively). Hence, the filamentation and the optical breakdown in the water-glycerol mixture must happen at lower laser peak powers (i.e., fluences, because the laser pulse duration and the focusing conditions were the same), than in water. Thus, we suppose that the decrease in the hardness in the figure~\ref{AllTogether}(d) is due to onset of the incident laser beam filamentation for fluences above approximately 3000~$J/cm^2$. 

\begin{figure}
\begin{center}
\includegraphics[width=8cm]{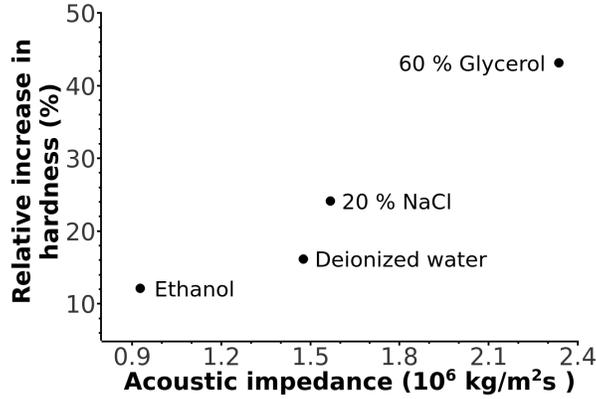}
\end{center}
\caption{Comparison of maximum relative increase in surface hardness of stainless-steels laser peened in different media of confinement at different respective acoustic impedances. The points corresponding to Ethanol and 20\% NaCl are for SUS316 and the other two points corresponding to Deionized water and 60\% Glycerol are for SUS316L}
\label{summary}
\end{figure}

To compare the effectiveness of all the confinement media used in the present study, we have plotted the relative increase in the maximal measured hardness of the stainless steels as a function of the acoustic impedance of the confinement medium, which is shown in figure~\ref{summary}. Since acoustic impedance is a product of the density and corresponding speed of sound in that medium it is therefore a very relevant parameter for the confinement effects of the medium. As can be seen in figure~\ref{summary}, the relative increase in the hardness rises monotonically with the acoustic impedance in the range of this parameter used in the present study. This is plausible because if the confinement medium is denser and shock wave travels faster in that the resulting containment of the shock energy and its delivery to the target metal will be more effective. Here one may argue that if the speed of sound is higher in a medium, the total duration of the shock exerted on the target might be shorter. But as we know the total duration of the shock applied on the target metal also depends on the thickness of the confinement medium and we expect to find an optimum duration of the shock that would result in the maximum peening effect in the target metal.

\section{Conclusions}
Two important inferences can be drawn from this experimental study on the fundamental role of the confinement media used during the LSP in the ultra-short laser pulse durations regime. First, in this ultra-short pulse duration regime of LSP, the non-linear mechanisms like self-focusing, filamentation, plasma breakdown and bubble formation at high repetition rate in the confinement media can significantly attenuate the input energy of the laser pulses. It is therefore of paramount importance to characterize these adverse effects comprehensively so that their influence is minimized while carrying out the LSP in ps - fs pulse regime. Second, it is found that the resulting surface hardness of the peened metal as a function of laser fluence depends significantly on the confinement media and the relative increase in the hardness increases monotonically with the acoustic impedance of the liquid of the confinement medium. Since LSP is a technology of industrial potential, it is imperative to accurately and comprehensively establish the role of different confinement media in this technology. Further experiments are underway to get deeper insight into the role of confinement media in the ultra-short pulsed laser shock peening.

\section*{ACKNOWLEDGMENTS}
Lalit M. Kukreja acknowledges financial support received from the Alexander von Humboldt Foundation, Germany under grant no. IND/1015352 for his visiting position at Applied Laser Technologies, Ruhr-Universit\"at Bochum in Germany. Jan Hoppius acknowledges financial support of DFG, Project GU 1075/8 in SPP 1839 Tailored Disorder.

\end{document}